\documentclass[longbib,twocolumn,twocolappendix]{aastex702}

\usepackage{amsmath}
\usepackage{subcaption}
\usepackage{booktabs}

\shorttitle{Vibration-driven crater relaxation}
\shortauthors{Narita and Katsuragi}

\begin{document}

\title{Scaling law for the diffusion coefficient in vibration-driven crater relaxation}

\author[0009-0004-5591-101X]{Hayato Narita}
\affiliation{Department of Earth and Space Science, The University of Osaka, 1-1 Machikaneyama, Toyonaka, 560-0043, Osaka, Japan}
\email{hayato.narita@ess.sci.osaka-u.ac.jp}

\author[0000-0002-4949-9389]{Hiroaki Katsuragi}
\affiliation{Department of Earth and Space Science, The University of Osaka, 1-1 Machikaneyama, Toyonaka, 560-0043, Osaka, Japan}
\email{katsuragi@ess.sci.osaka-u.ac.jp}

\begin{abstract}
Impact craters relax over long timescales, and this process is commonly described by a diffusion model. The diffusion coefficient determines the relaxation rate, and it has been estimated from observed crater shapes. This coefficient, however, represents the combined effect of several physical mechanisms. Its physical basis has not been revealed yet. In this study, we isolate the contribution of seismic vibration in a controlled experiment. Using a quasi-two-dimensional setup, we first confirm that a linear diffusion equation reproduces the crater relaxation. We then measure the diffusion coefficient and obtain the scaling form through systematic experiments. We find that the diffusion coefficient is proportional to the crater diameter. This dependence is not expected for simple diffusion. We interpret this dependence as arising because the thickness of the vibro-fluidized granular layer scales with the crater depth. This scaling relation is the main result of this study. We then apply it to the Moon with a parameterized model of impact-driven seismic spreading in the regolith layer. Integrating over the impact flux yields a macroscopic coefficient nearly proportional to crater diameter for shallow-layer-like spreading. Its magnitude depends on uncertain model parameters but is consistent with observations for plausible values. These results indicate that vibration is a physically plausible contributor to crater relaxation on the Moon.
\end{abstract}


\section{Introduction}\label{Introduction}
\begin{figure}
\centering
\begin{subfigure}{\linewidth}
  \includegraphics[width=\linewidth]{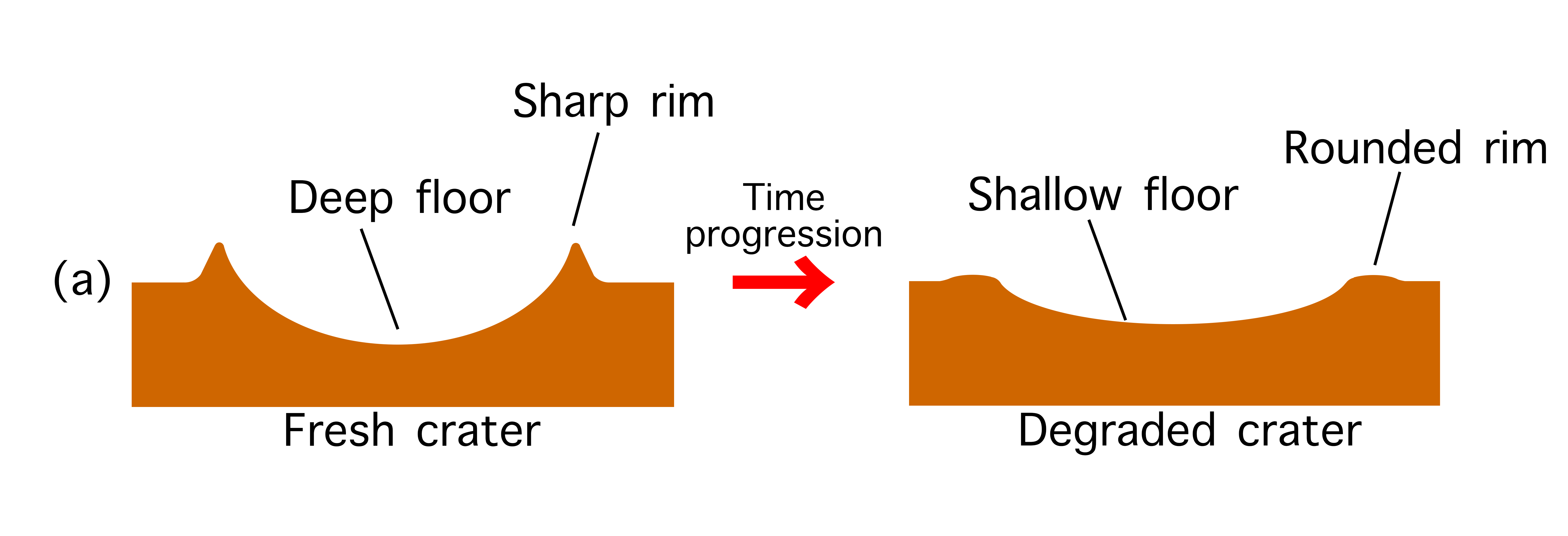}
  \phantomcaption
  \label{fig:relaxation_img}
\end{subfigure}

\vspace{-8mm}

\begin{subfigure}{\linewidth}
  \includegraphics[width=\linewidth]{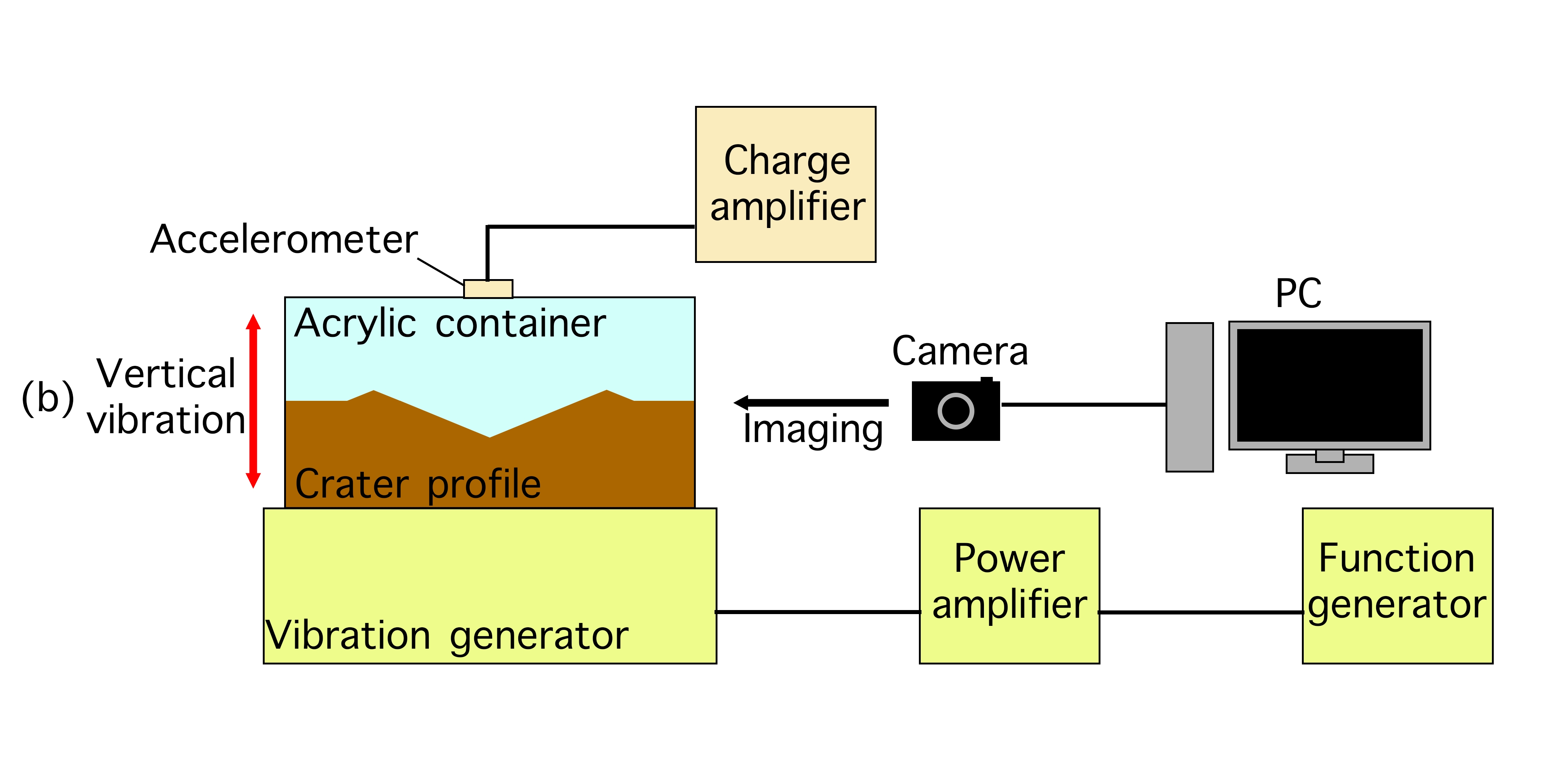}
  \phantomcaption
  \label{fig:setup}
\end{subfigure}

\vspace{-8mm}

\caption{(a)~Schematic images of a fresh crater and a relaxed (degraded) crater. Fresh craters have a deep floor, and a sharp and well-defined rim. Crater shape gradually relaxes with time. Degraded craters have a shallow floor, and a rounded and vague rim.
  (b)~Schematic images of the entire experimental setup. We made an initial crater profile in an acrylic container. The container was vertically vibrated by a vibration generator which is connected to a power amplifier and a function generator. The vibration acceleration was measured by an accelerometer and a charge amplifier. Then, we recorded videos of the relaxing crater shapes.
}
\label{fig:intro_method}
\end{figure}

There are numerous craters on the surface of solid celestial bodies. Impact cratering is a major process shaping planetary surfaces~\citep{Melosh1989,Melosh2011,Katsuragi2016}. Once a crater is formed by an impact, it relaxes gradually over a long time. As a result, planetary surfaces bear various generations of craters, ranging from fresh to old. The qualitative distinction is not difficult. A fresh crater has a sharp and well-defined rim. An old crater has a rounded and vague rim and a shallow floor~(Fig.~\ref{fig:relaxation_img}). The quantitative estimate from the shape of a single crater is, however, difficult. A diffusion equation has been used to model the shape evolution quantitatively. In this model, the diffusion coefficient sets the rate of crater relaxation. Several observational studies have estimated this coefficient from crater shapes~\citep[e.g.,][]{Fassett2014, Fassett2022}. The estimated coefficient represents the combined effect of several physical mechanisms. Its physical basis therefore has been left unsolved. In this study, we experimentally quantify the contribution of vibration to the diffusion coefficient. In addition, we quantitatively examine the consistency between our experimental result and model-based analysis of observational data.

Crater chronology is the standard tool to estimate the age of a planetary surface~\citep[e.g.,][]{Fassett2016}. It uses the size-frequency distribution of craters, because older surfaces accumulate more craters. For the Moon, returned samples fix the absolute ages of several sites~\citep[e.g.,][]{Tatsumoto1970, Wang2024}. These samples calibrate the chronology. Extending this calibration result, absolute ages can be assigned to arbitrary locations~\citep{Cohen2012, Yue2022}. This method works well in many cases. It can fail, however, where craters are too few or too many, or where secondary craters are present~\citep{Morota2008}. The chronology also gives an age averaged over an area, not a local age. If the formation age can be read from the shape of a single crater, the spatial resolution of dating improves. Estimating the age of an individual crater from its morphology requires a physically justified diffusion coefficient. Without such a coefficient, crater relaxation models remain largely empirical. 

We restrict the scope to simple craters. We do not treat craters with complex features such as floor-fractured craters~\citep[e.g.,][]{Salem2022}. We also exclude craters on icy bodies~\citep[e.g.,][]{Becq2025, Pamerleau2026}, because their relaxation mechanism is completely different. We mainly consider crater relaxation in a granular (regolith) layer.

Several processes drive crater relaxation. These include seismic vibration~\citep{Richardson2004,Richardson2005,Richardson2020}, ejecta deposition~\citep{Minton2019}, and small-scale impacts~\citep{Soderblom1970}. The relative importance of these processes depends on the size of the body. Seismic vibration is thought to dominate on small bodies such as asteroids. Ejecta deposition and other processes may become more important on larger bodies such as the Moon. The observed diffusion coefficient combines all of these contributions. To understand this coefficient, the contribution of each process should be evaluated separately. We focus on vibration first, because it is well suited to a controlled experiment. Ejecta deposition, in contrast, carries large uncertainty and is better treated numerically, as in \citet{Minton2019}. Once the vibration contribution is known, it can be compared with the other processes. We also note that relaxation by small-scale impacts shows behavior similar to vibration-driven relaxation of a granular heap~\citep{Omura2021}. The experimental approach to estimating the diffusion coefficient is therefore central to understanding planetary surface terrain development.

In this study, we measure the vibration-driven diffusion coefficient in a quasi-two-dimensional experiment. Although the actual crater relaxation occurs in three-dimensional space, a quasi-two-dimensional setup allows us to derive the essential physics with better measurement precision. The scaling form obtained here is dimensionless, which motivates its extrapolation to other geometries and gravity conditions, although this remains to be tested. We obtain its dependence on the vibration strength, the crater size, the grain size, and the friction, as a scaling relation. We then apply the result to the Moon, where the observed coefficient is well constrained~\citep{Fassett2022}. The resulting vibration model reproduces the observed near-linear diameter dependence under certain conditions and can match the observed magnitude for plausible model parameters. This indicates that vibration is a physically plausible contributor to crater relaxation, even on a large body.

\section{Method}\label{Method}
We constructed the experimental setup shown in Fig.~\ref{fig:setup}. For the sake of simplicity and measurement convenience, we conducted experiments in a quasi-two-dimensional system. The quasi-two-dimensional container was made of acrylic plates, with inner dimensions of $L=150$~mm in width, 100~mm in height, and 10~mm in depth. Glass beads with a reference grain size of 0.4~mm (AS-ONE Corp. BZ04) were poured into the container up to a height of 50~mm. After preparing the horizontally flat surface, the initial crater shape was created by pressing a wooden mold into the glass bead layer. The mold had the shape of an inverted isosceles-triangular cross-section with a base of 100~mm and a height of 20~mm as the reference geometry. This shape was selected because the diameter-to-depth ratio of fresh lunar craters is of order $5{:}1$~\citep{Pike1974, Pike1977}. The fixed ratio is a controlled experimental choice and is not intended to represent all lunar crater sizes. The crater diameter $D$ was varied by changing the pushing depth of the mold.

The container with the initial crater shape was then vertically vibrated by a vibration generator (EMIC, 513-B/A). A waveform generated by a function generator (N/F, WF1974) was sent to a power amplifier (EMIC, 374-A), which drove the vibration generator. We used sinusoidal waves in this experiment. During vibration, vibration acceleration was measured by an accelerometer attached to the container and connected to a charge amplifier. The deformation of the crater during vibration was recorded using a camera (OMRON SENTECH, STC-MCCM401U3V). The spatial resolution of the images was 0.20~mm/pixel, and the temporal resolution was 60~fps. The acquired image size was 1,100 $\times$ 700 pixels. Recording was continued for 10~s, which is sufficient to see the relaxation in our setup. The images recorded during the experiments were binarized. Then, the surface profiles of the craters were extracted for analysis. 

In this study, the control parameters are the peak acceleration $a$ and frequency of the vibration $f$, the initial crater diameter $D$, and the grain size $D_\mathrm{g}$ and shape of the particles (spherical glass beads or irregularly shaped Toyoura sand). The representative grain size of Toyoura sand is 0.23~mm. For Toyoura sand, the initial crater shape was prepared by fixing the mold in place, pouring sand around it, and then removing the mold, in order to retain a stable crater shape. The bulk friction coefficient $\mu$ of the glass beads and Toyoura sand was measured from the angle of repose as $\mu = \tan \theta_\mathrm{r}$, where $\theta_\mathrm{r}$ is the angle of repose. The measured values are 0.48 and 0.62, respectively for glass beads and Toyoura sand. For each experimental condition, five independent experiments were conducted, and the data used in the analysis represent the average values of the five trials with the error bars indicating the standard errors.

\begin{table}[t]
\centering
\caption{Parameters varied in our experiments and their corresponding ranges.}
\label{tab:parameters}
\begin{tabular}{ll}
\hline
Parameter & Range \\
\hline
Dimensionless peak acceleration ($\Gamma$) & 0.3--7 \\
Frequency ($f$) & 40--200 Hz \\
Initial crater diameter ($D$) & 0.05--0.1 m \\
Particle size ($D_\mathrm{g}$) & 0.2--2.0 mm \\
\hline
\end{tabular}
\end{table}

\section{Results}\label{Results}
\begin{figure*}
  \centering
  \includegraphics[width=\linewidth]{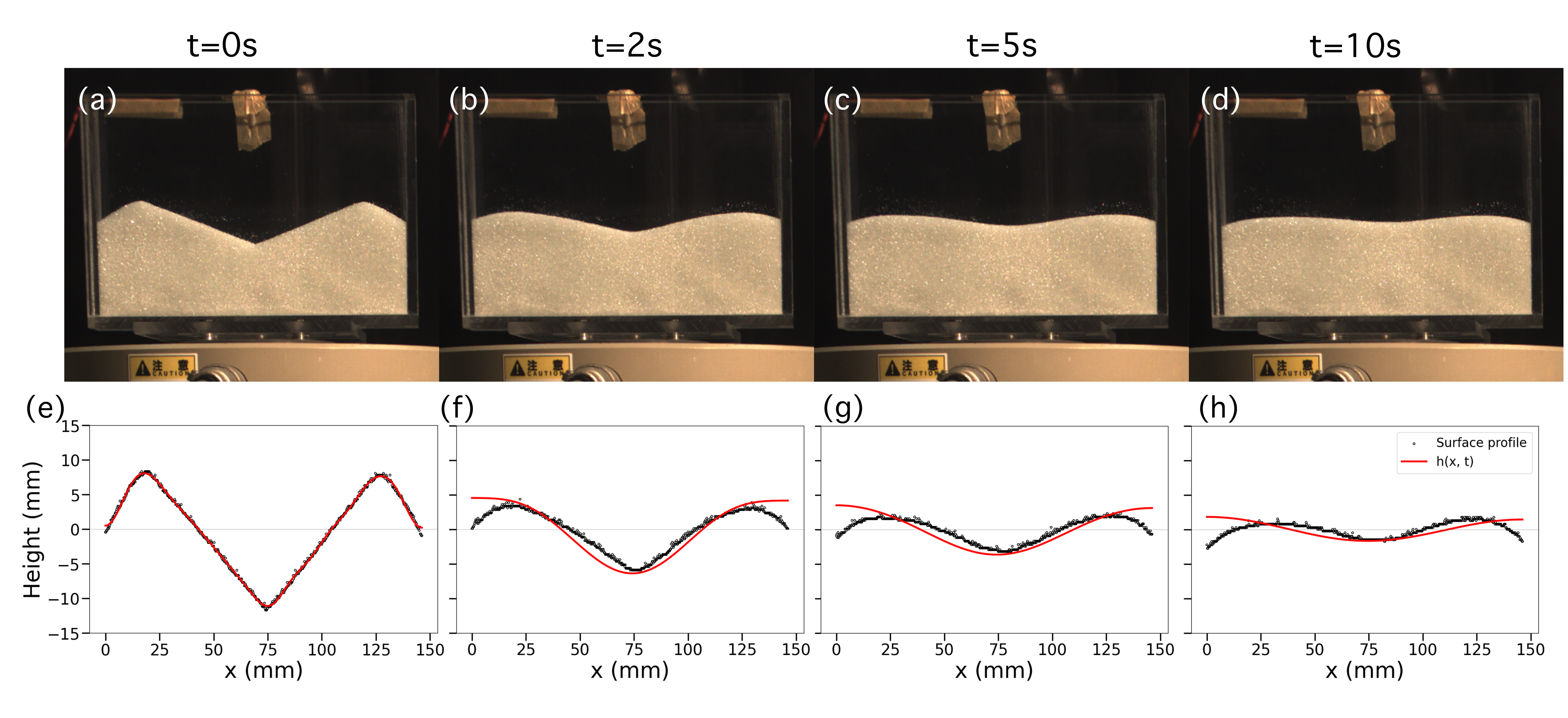}
  \caption{Crater relaxation example. Panels (a) -- (d) are the raw images of crater relaxation. The images (a) -- (d) show $t=0$, $2$, $5$, and $10$~s, respectively. The vibration acceleration of this experiment is $a=2g$, where $g=9.8$~m~s$^{-2}$ is gravitational acceleration. As shown in the difference between (a) and (b), short wavelength structures like rims are quickly relaxed at first. After $t=10$~s, the crater shape does not show significant deformation. The relaxation has almost saturated.
Images (e) -- (h) compare the experimental crater profile (black dots) and the model profile (red curve). The images (e) -- (h) display $t=0$, $2$, $5$, and $10$~s, respectively. From (e), the initial shape is well approximated by a Fourier expansion. As time proceeds, the experimental and theoretical crater profiles show some discrepancy. In the experiment, the crater profiles do not reach completely flat due to granular convection and other secondary effects.
}
  \label{fig:Raw_vs_Fitting}
\end{figure*}

Examples of the acquired images are shown in Fig.~\ref{fig:Raw_vs_Fitting}(a-d). The initial crater shape relaxes gradually, as expected. We evaluate the relaxation rate with a linear diffusion equation. A nonlinear diffusion form has been established for the relaxation of vibrated granular slopes~\citep{Roering1999, Tsuji2018}. The nonlinear form is needed to describe the full relaxation of a steep slope. The granular flux diverges as the slope angle approaches the angle of repose. A steep slope therefore relaxes very quickly. However, the long-lasting main part of the relaxation is governed by the linear term~\citep{Richardson2005}. The linear form is obtained by linearizing the nonlinear model at small slope angles. This linear model has been applied to quasi-two-dimensional granular flow~\citep{Xiao2017}. We use it here.

Therefore, we assume that the height of the profile $h$ at horizontal position $x$ and time $t$ obeys the linear diffusion equation,
\begin{equation}
\frac{\partial h(x,t)}{\partial t}=\kappa\frac{\partial^2h(x,t)}{\partial x^2},
\label{eq:diffusion_equation}
\end{equation}
where $\kappa$ is the diffusion coefficient. The linear diffusion equation has an analytic solution, 
\begin{equation}
h(x,t)=\sum_{n=0}^{\infty} C_n  \cos
\left(\frac{n\pi}{L}x
\right)\exp
\left[-\kappa
\left(\frac{n\pi}{L}
\right)^2t
\right],
\label{eq:fourier_expansion}
\end{equation}
where $C_n$ and $L$ are the Fourier coefficients and the width of the container, respectively. The boundary condition of no inflow or outflow at either end ($\partial h/\partial x =0$) is applied. In this study, the crater profiles are approximated by $20$th-order Fourier series ($0\leq n \leq 20$). 

First, the initial crater profile is approximated by the form of Eq.~(\ref{eq:fourier_expansion}), i.e., $C_0$, $C_1$, $\dots$, $C_{20}$ are obtained by the fitting. After turning on the vibration at $t=0$, the temporal development of the crater profile can be computed using Eq.~(\ref{eq:fourier_expansion}). This analytically estimated profile is compared with the experimental data. Then, the value of diffusion coefficient $\kappa$ can be estimated by least-squares fitting as shown in Fig.~\ref{fig:Raw_vs_Fitting}(e-h). The discrepancies between the experimental and analytical profiles near both ends probably come from the side wall effect. However, the model successfully reproduces the principal features of the crater relaxation. In this fitting, the diffusion coefficient $\kappa$ is the sole fitting parameter.

\begin{figure}
  \centering
  \includegraphics[width=\linewidth]{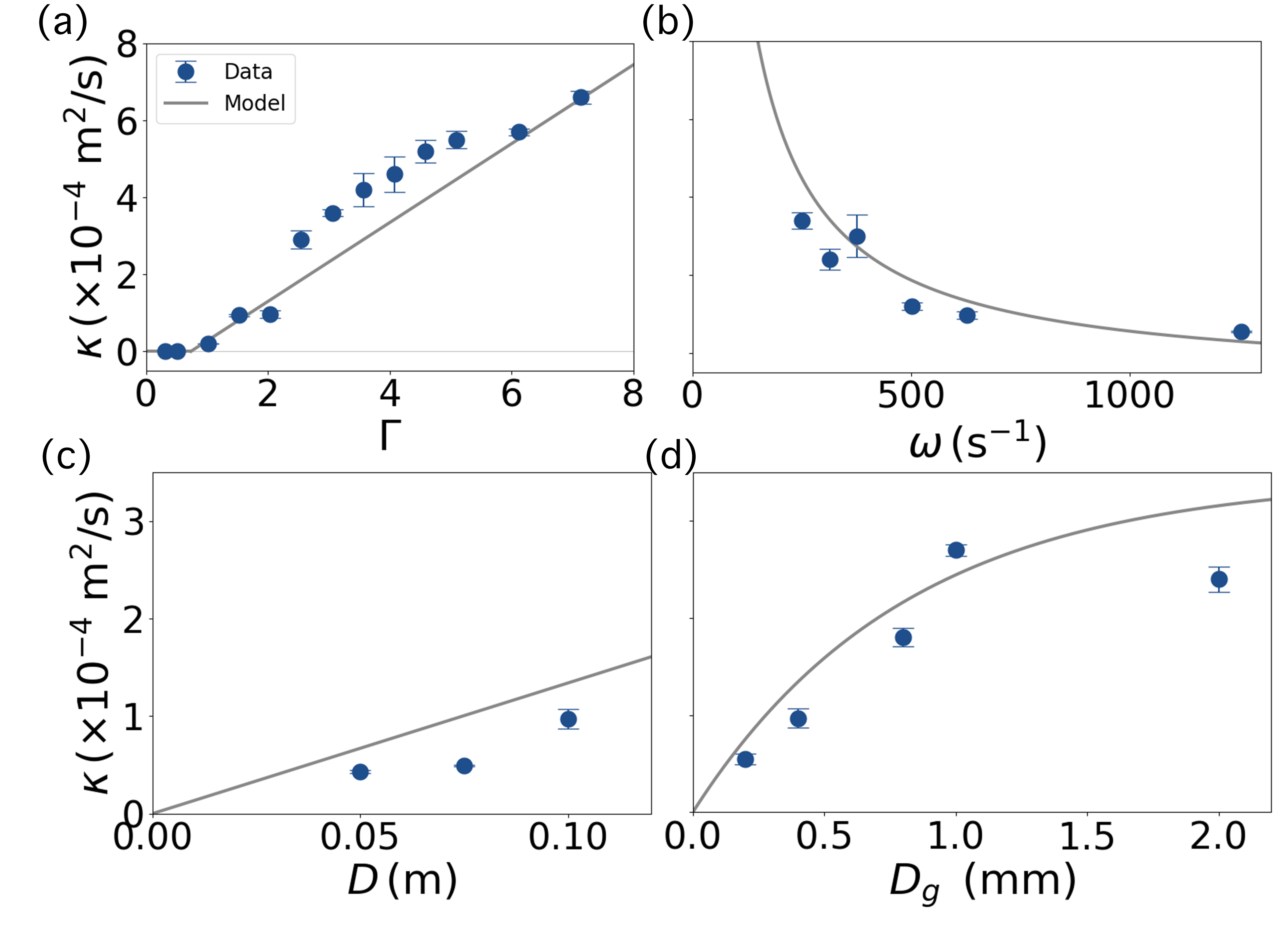}
  \caption{The parameter dependencies of $\kappa$. The relation between $\kappa$ and (a)~the vibration acceleration $\Gamma$, (b)~the vibration angular frequency $\omega$, (c)~the initial crater diameter $D$, and (d)~the grain size $D_\mathrm{g}$ is shown. In each plot, blue symbols are experimental data, and the gray curves are the fit to the model of Eq.~(\ref{eq:our_model}) using all data (Fig.~\ref{fig:Alldata}). The error bars in these plots originate from the experimental uncertainties (standard errors) obtained from the five repeated experiments.
}
\label{fig:param-k}
\end{figure}

Next, we examine the parameter dependence of $\kappa$. We performed a set of experiments by varying control parameters as shown in Table~\ref{tab:parameters}. The measured parameter dependence of $\kappa$ is plotted in Fig.~\ref{fig:param-k}. To characterize the strength of vibration, we use the dimensionless peak acceleration, $\Gamma = a/g$, where $a$ and $g$ are acceleration amplitude and gravitational acceleration, respectively. As presented in~Fig.~\ref{fig:param-k}, $\kappa$ is an increasing function of vibration acceleration $\Gamma$, initial crater diameter $D$, and constituent grain size $D_\mathrm{g}$, while it shows a decreasing trend as the vibration angular frequency $\omega=2\pi f$ increases. From these experimental results and the previous form of the diffusion coefficient~\citep{Tsuji2018}, we arrive at the following functional form
\begin{equation}
  \kappa = \begin{cases} \dfrac{\alpha}{\mu^2}\left(v-c_c\sqrt{gD_\mathrm{g}}\right)Df(D_\mathrm{g}) & \left(\dfrac{v}{\sqrt{gD_\mathrm{g}}} > c_c\right) \vspace{3mm} \\ 0 & \left(\dfrac{v}{\sqrt{gD_\mathrm{g}}} \le c_c\right). \end{cases}
\label{eq:our_model}
\end{equation}
Here, the function $f(D_\mathrm{g})$ is given by $f(D_\mathrm{g}) = 1 - \exp\left(-\frac{D_\mathrm{g}}{D^*_\mathrm{g}} \right)$. The quantity $v=a/\omega=\Gamma g/\omega$ is the peak vibration velocity, and $c_c$ is a dimensionless onset threshold, so that $v_c=c_c\sqrt{gD_\mathrm{g}}$. The fitting parameters are $\alpha$, $c_c$, and $D^*_\mathrm{g}$. The exponential saturation form of $D_\mathrm{g}$ was chosen as the simplest empirical function that reproduces the observed asymptotic behavior. In this experimental series, the reference conditions are set as $\Gamma=2$, $\omega=200 \pi$~s$^{-1}$, $D=0.1$~m, and $D_\mathrm{g}=0.4$~mm, and only one parameter is varied at a time to examine the parameter dependence of diffusion coefficient. 

\begin{figure}
  \centering
  \includegraphics[width=\linewidth]{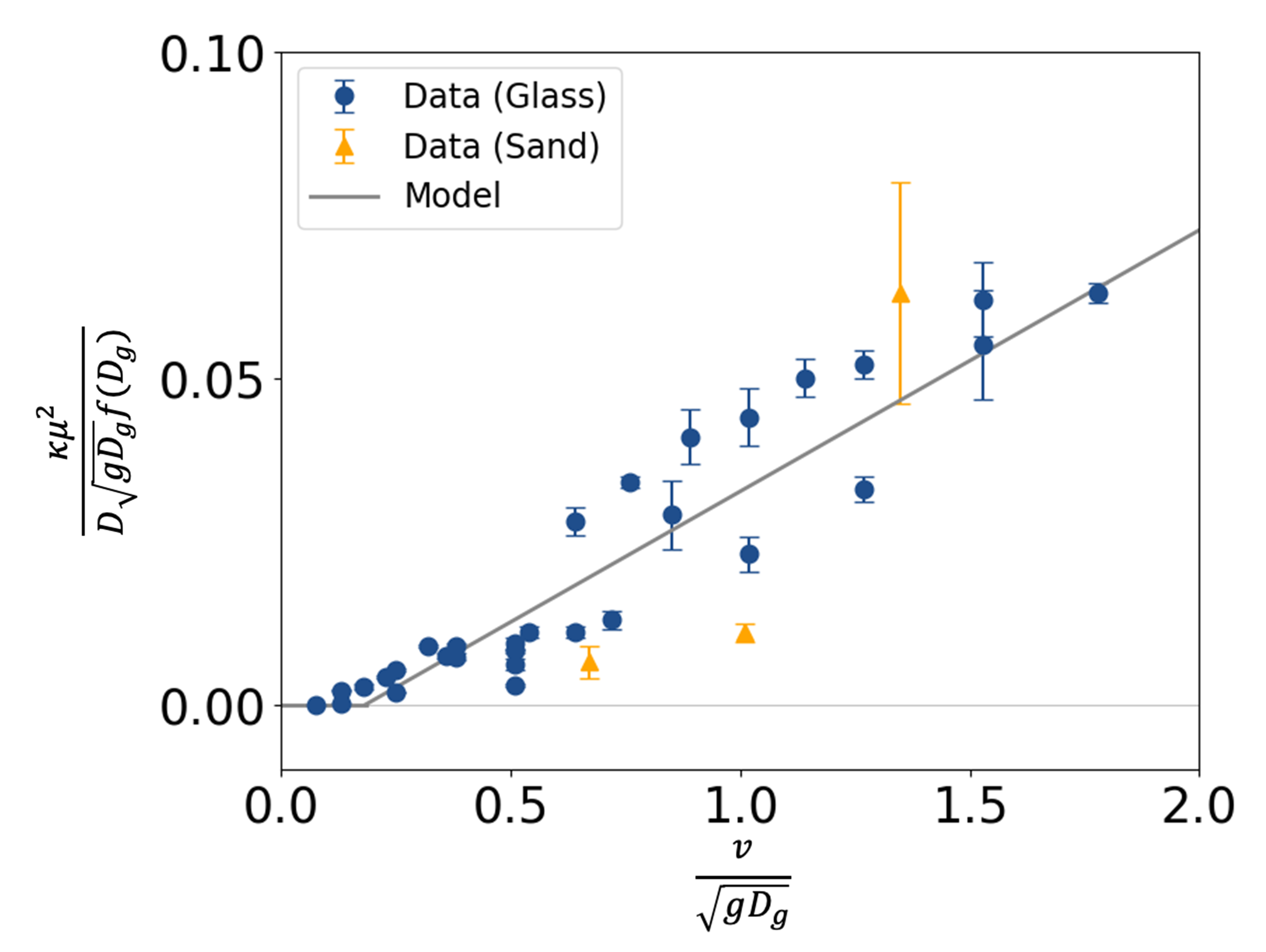}
  \caption{All data show a reasonable collapse to the scaling form of Eq.~(\ref{eq:our_model}). Blue circular symbols are experimental data of glass beads. Orange triangular symbols indicate the experimental data of Toyoura sand. The gray line is the fit to the scaling form of Eq.~(\ref{eq:our_model}). Both the horizontal and vertical axes are nondimensionalized. From the fitting, the dimensionless onset threshold $c_c=0.185$ is obtained. The error bars in these plots originate from the experimental uncertainties (standard errors) obtained from the five repeated experiments.
}
  \label{fig:Alldata}
\end{figure}

On the basis of the analysis so far, all the experimental data should align with the unified scaling relation of Eq.~(\ref{eq:our_model}). To validate the scaling form, dimensionless plot $\frac{\kappa \mu^2}{D\sqrt{g D_\mathrm{g}} f(D_\mathrm{g})}$ vs. $\frac{v}{\sqrt{gD_\mathrm{g}}}$ is plotted in Fig.~\ref{fig:Alldata}. One can confirm the data collapse to the master curve. From the fitting of all experimental data to the scaling form of Eq.~(\ref{eq:our_model}), the values of fitting parameters are obtained as $\alpha=4.0 \times 10^{-2}$, $c_c=0.185$, and $D_\mathrm{g}^*=8.12 \times 10^{-4}$~m. For reference, $c_c=0.185$ corresponds to $v_c=1.16\times10^{-2}$~m~s$^{-1}$ at $D_\mathrm{g}=0.4$~mm under Earth gravity. Although some scatter is present, the model given by Eq.~(\ref{eq:our_model}) successfully captures the overall trends of the experimental results. In this plot, Toyoura sand results are also plotted. These data are consistent with the assumed friction dependence $\kappa\propto \mu^{-2}$. Note that the $\kappa \propto \mu^{-2}$ dependence is assumed following \citet{Tsuji2018}. The data in Fig.~\ref{fig:param-k}(c) fall systematically below the fitted line. We attribute this offset to a finite-size (wall-friction) effect of the container. A larger container is needed to test this explanation and the $D$ scaling directly. 

\section{Discussion}\label{Discussion}
The scaling form, Eq.~(\ref{eq:our_model}), is the main experimental result of this study. In the following, we consider its meaning and apply this scaling to actual lunar craters. This application requires several additional assumptions, as described below.

\subsection{Meaning of the scaling of $\kappa$}
The obtained scaling of the diffusion coefficient contains a dependence on the crater diameter $D$. In classical linear diffusion, the diffusion coefficient is a local transport property of the medium. It does not depend on the size of the relaxing feature. In our result, however, $\kappa$ is proportional to $D$. This is not expected from simple diffusion. The physical origin of this dependence is examined below.

The $D$ dependence of $\kappa$ becomes clear when compared with previous granular experiments. In vibrated granular slopes, the local height $h$ is the characteristic length scale for $\kappa$~\citep{Tsuji2018,Tsuji2019}. This is because the granular flux driven by vibration is proportional to the height of the granular layer. In a crater, the same logic applies. The regolith layer is thick enough for the vibro-fluidized layer to develop. Because the diameter-depth ratio of the initial crater is fixed in this experiment, the crater depth is proportional to $D$. Under these conditions, we interpret the fluidized-layer thickness as being approximately proportional to $D$. Therefore, $D$ plays the same role as $h$ in the slope case. It enters the scaling as a measure of the mobile-layer thickness. Although the crater depth could be used as the length scale instead, $D$ is employed because it is easier to measure in observations. We also neglect temporal changes in the crater diameter and the characteristic mobile-layer thickness when interpreting the scaling. The rest of the parameter dependence matches the linearized coefficient for vibrated granular slopes~\citep{Tsuji2018,Tsuji2019}. The remaining difference is the finite dimensionless onset threshold $c_c$, which likely reflects a resolution issue in the previous studies. Because those studies focused on steep slopes, the onset of relaxation at small vibration velocities was not well resolved.

The $D$-dependent diffusion coefficient also appears in lunar observations. A study based on crater lifetimes reported $\kappa_\mathrm{obs} \propto D^{\xi}$, with $\xi= 0.87$ or $1.05$~\citep{Fassett2022}, where $\kappa_\mathrm{obs}$ is the observation-based diffusion coefficient. The observed coefficient is therefore close to proportional to $D$. To carefully compare the $D$ dependence of $\kappa$ and $\kappa_\mathrm{obs}$, the impact flux must also be considered. Therefore, in the next subsection, we estimate the macroscopic diffusion coefficient $\kappa_\mathrm{macro}$ by integrating the impact frequency model.

\subsection{Quantitative comparison with lunar observations}\label{Model_consistency_with_previous_study}
We estimate the macroscopic diffusion coefficient $\kappa_\mathrm{macro}$ for the Moon.
Here, $\kappa_\mathrm{macro}$ is the time-averaged coefficient produced by repeated impact-induced vibration events.
We obtain it by integrating the experimental single-event scaling, Eq.~(\ref{eq:our_model}), over the lunar impact flux.
We compare the result with the estimate of \citet{Fassett2022} in terms of both the $D$ dependence and the absolute magnitude.
The parameters used for the numerical evaluation are summarized in Table~\ref{tab:lunar_params}, and the full derivation is given in Appendix~\ref{app:kappa_macro}.
For the scaling analysis below, the parameters in Table~\ref{tab:lunar_params} and the experimental fit parameters in Section~\ref{Results} are treated as effective constants over the impactor-size range that dominates the integral and as independent of $D$ unless otherwise stated.
In particular, the two spatial factors introduced below are uncertain in value, but the fiducial model assumes that they have no additional systematic dependence on $D$.

An impactor of diameter $\delta$ delivers kinetic energy $E_\mathrm{imp}=\pi\rho_\mathrm{imp}v_\mathrm{imp}^2\delta^3/12$, where $\rho_\mathrm{imp}$ and $v_\mathrm{imp}$ are the impactor density and impact velocity, respectively.
A fraction $\eta$ of this energy is converted into seismic vibration energy~\citep{Richardson2005}.
The seismic energy density is written as $\epsilon_s=\rho v^2/2=\eta E_\mathrm{imp}/V_\mathrm{eff}$, where $\rho$ is the regolith bulk density and $V_\mathrm{eff}$ is the effective volume relevant to the vibration experienced by the crater.
The regolith--bedrock structure and the strongly scattered, weakly attenuated character of lunar seismic waves~\citep{Stopar2017,Horz1991,Nunn2020} motivate a shallow-layer-like spreading model.
For the fiducial case, we parameterize the effective volume as
\begin{equation}
V_\mathrm{eff}=C_sD^2h_\mathrm{reg},
\label{eq:V_main}
\end{equation}
where $h_\mathrm{reg}$ is the regolith-layer thickness and $C_s$ is a dimensionless seismic-spreading factor.
The factor $C_s$ represents lateral spreading, attenuation, leakage, and spatial nonuniformity of the vibration field.
It is not assumed to define a sharp boundary or to be of order unity.
For fixed material parameters, Eq.~(\ref{eq:V_main}) gives $v\propto(\eta/C_s)^{1/2}\delta^{3/2}/D$.
The detailed derivation is given in Appendix~\ref{sec:V} and Appendix~\ref{sec:velocity}.

The spatial range of impacts that can contribute to relaxation is parameterized independently.
We write the effective impact-collection area as
\begin{equation}
A_i=C_iD^2,
\label{eq:Ai_main}
\end{equation}
where $C_i$ is a dimensionless impact-collection factor.
The factor $C_i$ represents the spatial range and weighting of impact locations capable of relaxing the crater.
Thus, $C_i$ and $C_s$ describe different parts of the model and need not be equal. 
In the present model, $C_i$ determines the effective rate of impacts that contribute to the relaxation of a given crater. On the other hand, $C_s$ determines how the seismic energy from each contributing impact is spread before producing the vibration experienced by the crater. 
Thus, an impact contributing through $A_i$ supplies the kinetic energy $E_\mathrm{imp}$. A fraction $\eta$ of this energy is converted into seismic energy and distributed according to $V_\mathrm{eff}$. Namely, $C_i$ and $C_s$ indicate how many impacts contribute and how strongly each contributing impact shakes the crater, respectively.
For a cumulative impactor number distribution $N(>\delta)\propto\delta^{-\beta}$, where $\beta$ is the cumulative size-distribution exponent, the differential local impact rate scales as $d\lambda_\mathrm{local}/d\delta\propto C_iD^2\delta^{-(\beta+1)}$ (see Appendix~\ref{sec:rate} for the details including the definition of $\lambda_\mathrm{local}$ etc.).

The experimental onset condition is $v_c=c_c\sqrt{g_mD_\mathrm{g}}$ for lunar gravity $g_m$ and lunar regolith grain size $D_\mathrm{g}$.
We define $\delta_\mathrm{min}$ as the minimum impactor diameter for which the vibration exceeds this onset.
Because $v\propto(\eta/C_s)^{1/2}\delta^{3/2}/D$, the onset condition gives $\delta_\mathrm{min}\propto(C_s/\eta)^{1/3}D^{2/3}$ (Appendix~\ref{sec:dmin}).
We also introduce an upper cutoff $\delta_\mathrm{max}=C_\mathrm{max}D$, with $C_\mathrm{max}=10^{-2}$, to exclude impacts large enough to overwrite the crater (Appendix~\ref{sec:dmax}).
We denote by $\kappa(\delta,D)$ the diffusion coefficient produced by a single impactor of diameter $\delta$, and by $T_\mathrm{dur}$ the effective duration of vibration following one impact.
The macroscopic diffusion coefficient is then
\begin{equation}
\kappa_\mathrm{macro}(D)=T_\mathrm{dur}\int_{\delta_\mathrm{min}}^{\delta_\mathrm{max}}\kappa(\delta,D)\frac{d\lambda_\mathrm{local}}{d\delta}\,d\delta.
\label{eq:kappa_macro_integral_main}
\end{equation}
The full expressions entering Eq.~(\ref{eq:kappa_macro_integral_main}) are given in Appendix~\ref{sec:derivation}.
For $\beta>3/2$, the integral is dominated by impactors near $\delta_\mathrm{min}$.
Combining the above scalings gives
\begin{equation}
\kappa_\mathrm{macro} \propto C_i\left(\frac{\eta}{C_s}\right)^{\beta/3}D^{3-(2/3)\beta}.
\label{eq:kappa_macro_scaling}
\end{equation}
Thus, $\eta$, $C_s$, and $C_i$ determine the normalization, while the crater-diameter exponent is independent of their values as long as $C_s$ and $C_i$ have no additional $D$ dependence.
The algebra leading to Eq.~(\ref{eq:kappa_macro_scaling}) is shown in Appendix~\ref{sec:integral}.
The Brown-calibrated value $\beta\simeq2.7$~\citep{Brown2002} gives $\kappa_\mathrm{macro}\propto D^{1.2}$, while $\beta\simeq3.0$--$3.1$ gives $D^{0.93}$--$D^{1.00}$.
Both are close to the near-linear dependence reported by \citet{Fassett2022}.

The shallow-layer assumption can also be generalized.
We write $V_\mathrm{eff}=C_sD^qh_\mathrm{reg}^{3-q}$, where $q$ is the effective spreading exponent.
The fiducial shallow-layer case corresponds to $q=2$, while homogeneous three-dimensional spreading corresponds to $q=3$.
Keeping $A_i=C_iD^2$, the same derivation then gives $\kappa_\mathrm{macro}\propto D^{3-q\beta/3}$.
The observed crater-size exponent corresponds to $q\simeq1.9$--$2.4$ for $\beta\simeq2.7$--$3.1$ (Appendix~\ref{sec:comparison}).
Thus, the observations $\kappa_\mathrm{obs} \propto D$ favor shallow-layer-like effective spreading over homogeneous three-dimensional spreading within the present model.

\begin{table}[t]
\centering
\caption{Lunar and model parameters used in the numerical evaluation of $\kappa_\mathrm{macro}$. Experimental fit parameters are given in Section~\ref{Results}; a complete list with sources is given in Table~\ref{tab:params}.}
\label{tab:lunar_params}
\begin{tabular}{ll}
\hline
Parameter & Value \\
\hline
Lunar gravity ($g_m$) & 1.62~m~s$^{-2}$ \\
Regolith bulk density ($\rho$) & 1500~kg~m$^{-3}$ \\
Impactor density ($\rho_\mathrm{imp}$) & 3000~kg~m$^{-3}$ \\
Impact velocity ($v_\mathrm{imp}$) & 19.7~km~s$^{-1}$ \\
Seismic conversion efficiency ($\eta$) & $10^{-4}$ \\
Grain size ($D_\mathrm{g}$) & 57.5~$\mathrm{\mu m}$ \\
Friction coefficient ($\mu$) & 0.62 \\
Vibration duration ($T_\mathrm{dur}$) & 800~s \\
Regolith layer thickness ($h_\mathrm{reg}$) & 20~m \\
Effective spreading exponent ($q$) & 2 (fiducial) \\
Seismic-spreading factor ($C_s$) & $4\pi$ (fiducial) \\
Impact-collection factor ($C_i$) & $\pi/4$ (fiducial) \\
Impact-flux exponent ($\beta$) & 2.7 (fiducial) \\
Upper-cutoff factor ($C_\mathrm{max}$) & $10^{-2}$ \\
\hline
\end{tabular}
\end{table}

We evaluate the absolute magnitude using the fiducial values in Table~\ref{tab:lunar_params}.
For a concrete numerical normalization, we use $C_{s,0}=4\pi$ and $C_{i,0}=\pi/4$.
These reference values correspond to a radius-$2D$ spreading area and the crater footprint, respectively.
They are not treated as calibrated physical boundaries.
The detailed numerical evaluation is given in Appendix~\ref{sec:numerical}.
With $\eta=10^{-4}$, the result at $D=100$~m is $\kappa_\mathrm{macro}\simeq1.7$~m$^2$~Myr$^{-1}$.
The equilibrium estimate of \citet{Fassett2022} is $0.28$~m$^2$~Myr$^{-1}$ at the same diameter.
The fiducial prediction is therefore larger by a factor of about $6$.

The three quantities $\eta$, $C_s$, and $C_i$ are degenerate in this normalization.
For convenience, Appendix~\ref{sec:sensitivity} defines the effective seismic-efficiency parameter
$\eta_\mathrm{eff}\equiv\eta(C_i/C_{i,0})^{3/\beta}(C_{s,0}/C_s)$.
Matching the observational value requires $\eta_\mathrm{eff}\simeq1.3\times10^{-5}$ for $\beta=2.7$.
If $C_i$ and $C_s$ are kept at their fiducial values, this is simply equivalent to $\eta\simeq1.3\times10^{-5}$, within the broad range considered for seismic efficiency~\citep{Richardson2005}.
Other combinations of $\eta$, $C_s$, and $C_i$ give the same normalization.
Therefore, the absolute comparison should be regarded as an order-of-magnitude consistency test rather than a parameter-free prediction.

Our experimentally obtained scaling, $\kappa \propto D$, is consistent with $\kappa_\mathrm{obs}\propto D^\xi$ with $\xi\simeq1$.
This agreement supports the idea that vibration-driven layer dynamics contributes to the observed relaxation.
The observed coefficient need not be explained solely by vibration.
The micrometeoroid bombardment effect, modeled by \citet{Soderblom1970} and applied to the Moon by \citet{Xie2017}, yields a similar near-linear dependence ($\kappa\propto D^{0.93}$), despite arising from a distinct physical mechanism.
However, microimpact-driven and vibration-driven slope relaxation follow the same transport law at the single-event level~\citep{Omura2021}.
This may explain the similarity between the two mechanisms.
A more complete partition among mechanisms is left for future work.

The purpose of the present model is not to provide a unique description of lunar seismic transport.
It is to test whether a shallow-layer-like vibration model can reproduce the observed scaling and whether its magnitude is compatible with plausible parameter values.

\subsection{Future prospects}\label{Future_prospects}
The present experiment establishes the scaling of the vibration-driven diffusion coefficient. Several extensions are natural next steps. The experiment is quasi-two-dimensional, while the real process is three-dimensional. Although the dimensionless form motivates extrapolation, a three-dimensional experiment is needed to test whether the same scaling survives a change of geometry. A larger container would also help estimate the wall effect noted in Fig.~\ref{fig:param-k}(c). The vibration conditions can also be varied, including direction, attenuation, and intermittency, to better match natural impacts. We used glass beads and Toyoura sand. Regolith simulants in a vacuum chamber would be more realistic. Controlling the gravitational acceleration, however, is not easy. We use relatively large grains, so air drag is negligible. Application under different gravity therefore remains an extrapolation. A further extension would replace the impact-collection factor $C_i$ and the seismic-spreading factor $C_s$ with a spatially distributed impact-forcing model that integrates the vibration field over impact distance. This would provide a direct connection to cumulative small-impact degradation models such as \citet{Soderblom1970}.

\section{Conclusion}\label{Conclusion}
In this study, we measured the vibration-driven relaxation of craters in a quasi-two-dimensional experiment. We analyzed the crater profiles with a linear diffusion equation. The scaling form of the diffusion coefficient was obtained through systematic experiments. The coefficient is proportional to the crater diameter. We interpret this dependence as arising because the effectively fluidized-layer thickness is set by the crater depth. The scaling form is consistent with previous granular-heap experiments. This scaling relation is the main experimental result of this study. We assumed that individual impact-induced relaxation events accumulate approximately independently. We then integrated this scaling over the lunar impact flux with a parameterized shallow-layer seismic-spreading model. The model reproduces the observed near-linear diameter dependence, while its absolute normalization depends on the seismic efficiency, seismic-spreading factor, and impact-collection factor. For plausible parameter values, the predicted magnitude is consistent with observations at the order-of-magnitude level. The observed diffusion coefficient combines several mechanisms. Our results provide a first quantitative estimate of the vibration part of the diffusion coefficient. They show that vibration is a physically plausible contributor to crater relaxation. This provides a physical basis for future models of planetary surface evolution.

\section*{Data availability}
Supporting datasets have been deposited in Zenodo, including raw data underlying the results, the corresponding graphs for each sample, and numerical data used to generate the graphs. The data are available at \dataset[10.5281/zenodo.21378375]{https://doi.org/10.5281/zenodo.21378375}.

\begin{acknowledgments}
This work was partially supported by JSPS KAKENHI Grant Number JP24H00196 and JST ERATO Grant Number JPMJER2401. 
During the preparation of this work, the authors used Claude (Anthropic) to improve the clarity and phrasing of the manuscript text, and to cross-check intermediate numerical calculations. After using this tool, the authors reviewed and verified all content as needed. The authors take full responsibility for the content of the published article.
\end{acknowledgments}

\appendix

\section{Full derivation of the macroscopic diffusion coefficient}
\label{app:kappa_macro}

This appendix gives the full derivation of the macroscopic diffusion coefficient $\kappa_\mathrm{macro}(D)$ introduced in Section~\ref{Model_consistency_with_previous_study}. The starting point is the experimentally obtained scaling of Eq.~(\ref{eq:our_model}) in the main text, which gives the diffusion coefficient for a steady vibration of velocity $v$. To obtain $\kappa_\mathrm{macro}(D)$, three steps are needed. First, we relate the vibration velocity $v$ to the impactor diameter $\delta$ and the crater diameter $D$. Second, we substitute this relation into Eq.~(\ref{eq:our_model}) to obtain the diffusion coefficient produced by a single impactor, $\kappa(\delta,D)$. Third, we integrate $\kappa(\delta,D)$ over the lunar impact-rate distribution to obtain $\kappa_\mathrm{macro}(D)$.

All numerical values used below are listed in Table~\ref{tab:params}.
\begin{table*}[t]
\centering
\caption{Parameter values used in this appendix.}
\label{tab:params}
\begin{tabular}{lll}
\toprule
Parameter & Value & Source \\
\midrule
$g_m$ & $1.62$~m~s$^{-2}$ & Lunar gravity \\
$\rho$ & $1500$~kg~m$^{-3}$ & Representative value \\
$\rho_\mathrm{imp}$ & $3000$~kg~m$^{-3}$ & Representative value \\
$v_\mathrm{imp}$ & $19.7$~km~s$^{-1}$ & \citet{LeFeuvre2011} \\
$\eta$ & $10^{-4}$ & Seismic conversion efficiency \citep{Richardson2005} \\
$\alpha$ & $4.0\times10^{-2}$ & Experimental fit (Section~\ref{Results}) \\
$c_c$ & $0.185$ & Dimensionless onset threshold (Section~\ref{Results}) \\
$v_{c,\mathrm{lunar}}$ & $1.8\times10^{-3}$~m~s$^{-1}$ & Eq.~(\ref{eq:vclunar}) \\
$\mu$ & $0.62$ & \citet{Nie2023} \\
$D_\mathrm{g}$ & $57.5$~$\mu$m & \citet{Wu2025} \\
$D_\mathrm{g}^*$ & $8.12\times10^{-4}$~m & Experimental fit (Section~\ref{Results}) \\
$T_\mathrm{dur}$ & $800$~s & \citet{Nunn2020}, \citet{Onodera2024} \\
$h_\mathrm{reg}$ & $20$~m & \citet{Stopar2017}, \citet{Horz1991} \\
$q$ & $2$ & Fiducial effective spreading exponent \\
$C_s$ & $4\pi$ & Fiducial seismic-spreading factor \\
$C_i$ & $\pi/4$ & Fiducial impact-collection factor \\
$C_\mathrm{max}$ & $10^{-2}$ & Adopted, see text \\
$\beta$ & $2.7$ & \citet{Brown2002}, Eq.~(\ref{eq:beta_def}) \\
$R_\mathrm{Moon}$ & $1737$~km & Lunar radius \\
$R_\mathrm{Earth}$ & $6371$~km & Earth radius \\
\bottomrule
\end{tabular}
\end{table*}

\subsection{Effective seismic spreading in the regolith layer}
\label{sec:V}

An impact converts a fraction $\eta$ of its kinetic energy into seismic vibration energy. \citet{Richardson2005} give the resulting seismic energy density $\epsilon_s$ as
\begin{equation}
  \epsilon_s = \frac{\rho a^2}{2\omega^2} = \frac{\rho v^2}{2} = \frac{\eta E_\mathrm{imp}}{V_\mathrm{eff}},
\label{eq:Richardson5}
\end{equation}
where $\rho$ is the target (regolith) bulk density, $a$ is the vibration acceleration, $\omega=2\pi f$ is the vibration angular frequency, $v$ is the peak vibration velocity, $E_\mathrm{imp}$ is the impact kinetic energy, and $V_\mathrm{eff}$ is an effective volume relevant to the vibration experienced by the crater.

On the Moon, a granular regolith layer of thickness $h_\mathrm{reg}$ overlies megaregolith or bedrock~\citep{Stopar2017}. The acoustic impedance contrast at this boundary, together with strong scattering and weak attenuation of lunar seismic waves~\citep{Nunn2020}, suggests shallow-layer-like spreading of vibration energy. Instead of imposing a sharp lateral boundary, we write
\begin{equation}
V_\mathrm{eff}=C_sD^2h_\mathrm{reg},
\label{eq:V}
\end{equation}
where $C_s$ is the dimensionless seismic-spreading factor introduced in Section~\ref{Model_consistency_with_previous_study}. It incorporates lateral spreading, attenuation, leakage, and spatial nonuniformity of the vibration field relevant to a crater of diameter $D$. No sharp lateral boundary or order-unity value is assumed for $C_s$. For fixed $h_\mathrm{reg}$, the vibration amplitude depends on $\eta$ and $C_s$ only through $\eta/C_s$. We use $C_{s,0}=4\pi$ as a fiducial normalization for the numerical comparison, not as a calibrated physical boundary.

We adopt $h_\mathrm{reg}=20$~m as a representative upper value~\citep{Stopar2017,Horz1991}. More generally, we write
\begin{equation}
V_\mathrm{eff}=C_sD^q h_\mathrm{reg}^{3-q},
\label{eq:Vq}
\end{equation}
where $q$ parameterizes the effective dimensionality of seismic spreading. The fiducial shallow-layer case is $q=2$, whereas homogeneous three-dimensional spreading corresponds to $q=3$. The consequences of varying $q$ are discussed in Appendix~\ref{sec:comparison}.

\subsection{Step-by-step derivation of $\kappa_\mathrm{macro}(D)$}
\label{sec:derivation}

\subsubsection{Seismic energy density and vibration velocity}
\label{sec:velocity}

An impactor of diameter $\delta$ delivers kinetic energy
\begin{equation}
E_\mathrm{imp}(\delta) = \frac{1}{2}m_\mathrm{imp}v_\mathrm{imp}^2 = \frac{\pi\rho_\mathrm{imp}v_\mathrm{imp}^2}{12}\,\delta^3 \equiv A_E\,\delta^3,
\label{eq:Eimp}
\end{equation}
where $m_\mathrm{imp}$, $\rho_\mathrm{imp}$, and $v_\mathrm{imp}$ are the impactor mass, density, and impact velocity, respectively. The density and velocity are treated as fixed representative values~(Table~\ref{tab:params}). For the fiducial $q=2$ case, substituting Eq.~(\ref{eq:Eimp}) and Eq.~(\ref{eq:V}) into Eq.~(\ref{eq:Richardson5}) gives the seismic energy density,
\begin{equation}
\epsilon_s(\delta,D) = \frac{\eta A_E\,\delta^3}{C_sD^2 h_\mathrm{reg}}.
\label{eq:eps}
\end{equation}
Solving Eq.~(\ref{eq:Richardson5}) for the vibration velocity $v=a/\omega$ then gives
\begin{equation}
v(\delta,D) = \sqrt{\frac{2\epsilon_s}{\rho}} = C_v\,\frac{\delta^{3/2}}{D}, \qquad
C_v \equiv \sqrt{\frac{2\eta A_E}{\rho C_s h_\mathrm{reg}}}.
\label{eq:v}
\end{equation}
Thus, $C_v$ depends on the uncertain spreading geometry and seismic conversion only through $\eta/C_s$. As shown in Appendix~\ref{sec:cancel}, $v_\mathrm{imp}$ and $\rho_\mathrm{imp}$ do not affect $\kappa_\mathrm{macro}$ for the Brown-calibrated flux normalization.

\subsubsection{Impactor-level diffusion coefficient}
\label{sec:kappa_impactor}

Substituting Eq.~(\ref{eq:v}) into Eq.~(\ref{eq:our_model}) gives the diffusion coefficient produced by a single impactor of diameter $\delta$,
\begin{equation}
\kappa(\delta,D) = \frac{\alpha f(D_\mathrm{g})}{\mu^2}\left(C_v\,\frac{\delta^{3/2}}{D} - v_{c,\mathrm{lunar}}\right)D.
\label{eq:kexp_D2}
\end{equation}

\subsubsection{Lower cutoff: relaxation-onset threshold}
\label{sec:dmin}

Eq.~(\ref{eq:our_model}) gives the onset criterion directly in dimensionless form: $v/\sqrt{gD_\mathrm{g}}=c_c$. We therefore define $\delta_\mathrm{min}$ by the corresponding lunar threshold $v(\delta_\mathrm{min},D)=v_{c,\mathrm{lunar}}$. From the experimental collapse, $c_c=0.185$. Rescaling this dimensionless threshold to lunar gravity $g_m=1.62$~m~s$^{-2}$ and the lunar regolith grain size $D_\mathrm{g}=57.5~\mu$m (Table~\ref{tab:params}) gives
\begin{equation}
v_{c,\mathrm{lunar}} = c_c\sqrt{g_m D_\mathrm{g}} = 1.8\times10^{-3}~\mathrm{m~s^{-1}}.
\label{eq:vclunar}
\end{equation}
Setting $v(\delta_\mathrm{min},D)=v_{c,\mathrm{lunar}}$ in Eq.~(\ref{eq:v}) defines the minimum effective impactor diameter,
\begin{equation}
\delta_\mathrm{min}(D) = \left(\frac{v_{c,\mathrm{lunar}}\,D}{C_v}\right)^{2/3}.
\label{eq:dmin}
\end{equation}
This gives $\delta_\mathrm{min}\propto D^{2/3}$. Note that this threshold does not depend on an assumed lunar vibration frequency.

\subsubsection{Upper cutoff: crater-overwriting condition}
\label{sec:dmax}

A very large impact would overwrite the crater of interest rather than perturbing it. To exclude such events, we set an upper cutoff
\begin{equation}
\delta_\mathrm{max} = C_\mathrm{max}\,D,
\label{eq:dmax}
\end{equation}
with $C_\mathrm{max}=10^{-2}$, a small fraction chosen so that the impactor does not produce a crater comparable to $D$. As shown below, the integral over $\delta$ is dominated by its lower limit, so the precise value of $C_\mathrm{max}$ has only a minor effect on the result.

\subsubsection{Local impact rate}
\label{sec:rate}

The number of impactors per year that can contribute to relaxation of a crater of diameter $D$, with diameter in the range $[\delta,\delta+d\delta]$, is denoted $d\lambda_\mathrm{local}/d\delta$. Rather than imposing a hard impact-point boundary, we parameterize the effective collection area as
\begin{equation}
A_i=C_iD^2,
\label{eq:Ai}
\end{equation}
where $C_i$ is the dimensionless impact-collection factor introduced in Section~\ref{Model_consistency_with_previous_study}. It represents the spatial range and weighting of impact points capable of relaxing the crater. 
It is conceptually distinct from $C_s$. The factor $C_i$ determines the effective rate of contributing impacts, whereas $C_s$ determines the seismic energy density, and hence the vibration amplitude, produced by each contributing impact.

\citet{Brown2002} give the cumulative annual number of bolide impacts on Earth with impact energy greater than $E$ as
\begin{equation}
\log_{10}N_E(>E) = 0.5677 - 0.90\log_{10}E,
\label{eq:brown}
\end{equation}
with $E$ in kilotons of TNT, $1$~kton~$=E_\mathrm{kton}\equiv4.184\times10^{12}$~J. Writing $E=E_\mathrm{imp}(\delta)=A_E\delta^3$ from Eq.~(\ref{eq:Eimp}) and converting Eq.~(\ref{eq:brown}) to a power law in $\delta$ gives
\begin{align}
N_\mathrm{Earth}(>\delta) &= C_0\,\delta^{-\beta}, \label{eq:NEarth}\\
C_0 &\equiv 10^{0.5677}\left(\frac{A_E}{E_\mathrm{kton}}\right)^{-0.9}, \label{eq:C0}\\
\beta &= 3\times0.90 = 2.7. \label{eq:beta_def}
\end{align}
The differential form is
\begin{equation}
\left|\frac{dN_\mathrm{Earth}}{d\delta}\right| = \beta C_0\,\delta^{-(\beta+1)}.
\label{eq:dNEarth}
\end{equation}
Scaling this global Earth rate to the Moon by surface area and multiplying by the effective collection area gives
\begin{align}
\frac{d\lambda_\mathrm{local}}{d\delta} &= \left|\frac{dN_\mathrm{Earth}}{d\delta}\right|\left(\frac{R_\mathrm{Moon}}{R_\mathrm{Earth}}\right)^2\frac{C_iD^2}{4\pi R_\mathrm{Moon}^2} \nonumber\\
&= C_\lambda D^2\delta^{-(\beta+1)}, \label{eq:dlambda}
\end{align}
with
\begin{equation}
C_\lambda\equiv\frac{\beta C_0C_i}{4\pi R_\mathrm{Earth}^2}.
\label{eq:Clambda}
\end{equation}
Here, $R_\mathrm{Moon}$ and $R_\mathrm{Earth}$ are the lunar and terrestrial radii, respectively. The fiducial choice $C_{i,0}=\pi/4$ corresponds to the crater footprint and recovers the numerical normalization used below. The Earth-to-Moon scaling uses the ratio of surface areas only. It does not include the difference in gravitational focusing between the two bodies. Including this difference would revise the normalization by a factor of order unity. The Brown power-law fit is calibrated for impactor diameters above about 1~m. The lower cutoff $\delta_\mathrm{min}$ used below (Appendix~\ref{sec:numerical}) is smaller than this value. The flux at $\delta_\mathrm{min}$ is therefore an extrapolation of Eq.~(\ref{eq:NEarth}). Possible changes in the size-distribution slope are represented by varying $\beta$.

\subsubsection{Duration of a single vibration event}
\label{sec:Tdur}

Each impact excites the regolith layer for a finite duration $T_\mathrm{dur}$, after which the vibration has decayed and no longer contributes to relaxation. On Earth, seismic energy decays quickly because of strong intrinsic attenuation. On the Moon, however, seismograms recorded during the Apollo missions show coda waves that persist far longer than their terrestrial counterparts, a consequence of the low attenuation and strong scattering in the highly fractured lunar crust \citep{Nunn2020}. \citet{Richardson2005} note a similar effect for small asteroids, where the whole body continues to vibrate after an impact because there is little material to absorb the energy. This long persistence of shaking is consistent with a shallow, strongly scattering vibration field in which seismic energy remains active near the surface rather than escaping promptly into deeper material. Based on Fig.~13 of \citet{Nunn2020}, we adopt $T_\mathrm{dur}=800$~s as a representative duration for a single impact-induced vibration event. Seismograms of meteoroid impacts on the Moon show a coda that lasts $20$~min or longer~\citep{Onodera2024}. Thus, this choice of $T_\mathrm{dur}$ is conservative. This persistence stems from weak intrinsic attenuation, an independently established property of the lunar crust, and does not rely on a sharp lateral boundary in the effective-volume parameterization of Appendix~\ref{sec:V}.

\subsubsection{Assembly and evaluation of the integral}
\label{sec:integral}

The macroscopic diffusion coefficient is obtained by integrating the impactor-level coefficient of Eq.~(\ref{eq:kexp_D2}), weighted by the local impact rate of Eq.~(\ref{eq:dlambda}), over impactor diameter, and multiplying by $T_\mathrm{dur}$:
\begin{equation}
\kappa_\mathrm{macro}(D) = T_\mathrm{dur}\int_{\delta_\mathrm{min}}^{\delta_\mathrm{max}}\kappa(\delta,D) \frac{d\lambda_\mathrm{local}}{d\delta}\,d\delta.
\label{eq:integral}
\end{equation}
This equation implicitly assumes linear superposition of individual impact-induced relaxation events. This assumption is analogous to cumulative treatments commonly used for repeated granular perturbations. 
By construction, $\kappa$ vanishes exactly at $\delta=\delta_\mathrm{min}$ (Appendix~\ref{sec:dmin}), so the full expression of Eq.~(\ref{eq:kexp_D2}) is kept with the lunar threshold $v_{c,\mathrm{lunar}}$, together with Eq.~(\ref{eq:dlambda}):
\begin{equation}
\begin{aligned}
\kappa\frac{d\lambda_\mathrm{local}}{d\delta} = {}& \frac{\alpha f(D_\mathrm{g})}{\mu^2}\,C_\lambda\,D^2 \\
&\times\left(C_v\,\delta^{1/2-\beta} - v_{c,\mathrm{lunar}}\,D\,\delta^{-(\beta+1)}\right).
\end{aligned}
\label{eq:integrand}
\end{equation}
Since $\beta\simeq2.7$, the exponents $1/2-\beta$ and $-(\beta+1)$ are both below $-1$, so both terms diverge as $\delta\to0$ and the integral of Eq.~(\ref{eq:integral}) is dominated by its lower limit,
\begin{equation}
\begin{aligned}
\int_{\delta_\mathrm{min}}^{\delta_\mathrm{max}}\delta^{1/2-\beta}\,d\delta &\simeq \frac{\delta_\mathrm{min}^{3/2-\beta}}{\beta-3/2}, \\
\int_{\delta_\mathrm{min}}^{\delta_\mathrm{max}}\delta^{-(\beta+1)}\,d\delta &\simeq \frac{\delta_\mathrm{min}^{-\beta}}{\beta}.
\end{aligned}
\label{eq:int_approx}
\end{equation}
Physically, this means that relaxation is driven mainly by the numerous small impactors just above the relaxation-onset threshold, rather than by rare large impacts. Substituting $\delta_\mathrm{min}\propto D^{2/3}$ from Eq.~(\ref{eq:dmin}), both integrals in Eq.~(\ref{eq:integral}) through Eq.~(\ref{eq:integrand}) scale identically with $D$:
\begin{equation}
\begin{aligned}
D^2\int_{\delta_\mathrm{min}}^{\delta_\mathrm{max}}\delta^{1/2-\beta}\,d\delta &\propto D^2\left(D^{2/3}\right)^{3/2-\beta} = D^{3-(2/3)\beta}, \\
D^2\int_{\delta_\mathrm{min}}^{\delta_\mathrm{max}}D\delta^{-(\beta+1)}\,d\delta &\propto D^3\left(D^{2/3}\right)^{-\beta} = D^{3-(2/3)\beta}.
\end{aligned}
\label{eq:int_D}
\end{equation}
Keeping the onset $v_{c,\mathrm{lunar}}$ term explicit therefore changes the numerical prefactor of $\kappa_\mathrm{macro}$ but not its $D$-dependence. In the same lower-limit-dominated approximation, the dependence on the seismic-spreading and impact-collection factors is written as
\begin{equation}
\kappa_\mathrm{macro}\propto C_i\left(\frac{\eta}{C_s}\right)^{\beta/3}D^{3-(2/3)\beta},
\label{eq:normalization_scaling}
\end{equation}
when the other parameters are held fixed. Thus, $C_i$, $C_s$, and $\eta$ are degenerate in the absolute normalization, whereas the $D$ exponent is unchanged. The Brown-calibrated value $\beta\simeq2.7$ gives $\kappa_\mathrm{macro}\propto D^{1.2}$. For $\beta\simeq3.0$--$3.1$, close to the production function of \citet{Neukum2001} used by \citet{Fassett2022}, this gives $\kappa_\mathrm{macro}\propto D^{0.93}$--$D^{1.00}$, consistent with the near-linear dependence reported there.

\subsection{Evaluation at $D=100$~m}
\label{sec:numerical}

We evaluate the specific value of $\kappa_\mathrm{macro}$ at $D=100$~m, a typical crater diameter for which \citet{Fassett2022} report an equilibrium-based estimate. Combining Eqs.~(\ref{eq:integrand}) and (\ref{eq:int_approx}) with the multiplicative factor $T_\mathrm{dur}$ of Eq.~(\ref{eq:integral}), the lower-limit-dominated approximation for $\kappa_\mathrm{macro}(D)$ is written as
\begin{equation}
\begin{aligned}
\kappa_\mathrm{macro}(D) \simeq {}& T_\mathrm{dur}\,\frac{\alpha f(D_\mathrm{g})}{\mu^2}\,C_\lambda\,D^2 \\
&\times\left(C_v\,\frac{\delta_\mathrm{min}(D)^{3/2-\beta}}{\beta-3/2}\right. \\
&\left. - v_{c,\mathrm{lunar}}\,D\,\frac{\delta_\mathrm{min}(D)^{-\beta}}{\beta}\right).
\end{aligned}
\label{eq:kappa_full}
\end{equation}
Every quantity in Eq.~(\ref{eq:kappa_full}) is either listed in Table~\ref{tab:params} or defined by a numbered equation above: $C_v$ by Eq.~(\ref{eq:v}), $C_\lambda$ by Eq.~(\ref{eq:Clambda}), $v_{c,\mathrm{lunar}}$ by Eq.~(\ref{eq:vclunar}), and $\delta_\mathrm{min}(D)$ by Eq.~(\ref{eq:dmin}). For the fiducial normalization $C_s=4\pi$ and $C_i=\pi/4$, and with $D$ and $\delta$ measured in meters, $C_v=12.7$~m$^{1/2}$~s$^{-1}$, $C_\lambda=1.6\times10^{-13}$~yr$^{-1}$~m$^{\beta-2}$, and $\delta_\mathrm{min}=0.058$~m at $D=100$~m (Eq.~(\ref{eq:dmin})). The two terms of Eq.~(\ref{eq:kappa_full}) yield $3.0$ and $1.3$~m$^2$~Myr$^{-1}$. Multiplying by $T_\mathrm{dur}$ in seconds cancels the per-second dimension of $\kappa$, leaving $\kappa_\mathrm{macro}$ in m$^2$~yr$^{-1}$, which is converted to m$^2$~Myr$^{-1}$ by multiplying by $10^6$. Substituting these into Eq.~(\ref{eq:kappa_full}) gives
\begin{equation}
\kappa_\mathrm{macro}(D=100~\mathrm{m}) \simeq 1.7~\mathrm{m^2\,Myr^{-1}}.
\label{eq:kappa100}
\end{equation}
This is the fiducial value quoted in Section~\ref{Model_consistency_with_previous_study}. It is larger than the equilibrium estimate of \citet{Fassett2022}, $\kappa_\mathrm{macro,F}(D=100~\mathrm{m})=3.1\times(0.1)^{1.05}=0.28$~m$^2$~Myr$^{-1}$, by about a factor of 6. Because $\eta$, $C_s$, and $C_i$ are not independently constrained by the absolute normalization, this difference should not be interpreted as a precision discrepancy.

\subsubsection{Cancellation of $v_\mathrm{imp}$ and $\rho_\mathrm{imp}$ in the present normalization}
\label{sec:cancel}

Within the present Brown-calibrated normalization, the value of $\kappa_\mathrm{macro}$ in Eq.~(\ref{eq:kappa100}) does not depend on $v_\mathrm{imp}$ or $\rho_\mathrm{imp}$. Both enter only through $A_E$ in Eq.~(\ref{eq:Eimp}), and $A_E\propto \rho_\mathrm{imp}v_\mathrm{imp}^2$. We track the power of $A_E$ through the full derivation.

$C_v$ scales as $A_E^{1/2}$, from Eq.~(\ref{eq:v}). $v_{c,\mathrm{lunar}}$ does not depend on $A_E$ (Eq.~(\ref{eq:vclunar})). $\delta_\mathrm{min}$ scales as $C_v^{-2/3}$, from Eq.~(\ref{eq:dmin}). Thus, $\delta_\mathrm{min}$ scales as $A_E^{-1/3}$. In Eq.~(\ref{eq:kappa_full}), the first term contains one direct factor of $C_v$ and one factor of $\delta_\mathrm{min}^{3/2-\beta}$, while the second term (independent of $C_v$) contains one factor of $\delta_\mathrm{min}^{-\beta}$. With $\beta=2.7$, the exponent $3/2-\beta$ equals $-6/5$. For the first term, combining these two routes gives a power of $C_v$ of $9/5$, hence a power of $A_E$ of $\frac95\times\frac12=\frac9{10}$. For the second term, $\delta_\mathrm{min}^{-\beta}=\delta_\mathrm{min}^{-2.7}$ scales as $A_E^{-1/3\times(-2.7)}=A_E^{9/10}$, the same power. Both terms of $\kappa_\mathrm{macro}$ carry the same power of $A_E$.

Separately, $C_0$ in Eq.~(\ref{eq:NEarth}) scales as $A_E^{-9/10}$, since the Brown calibration of Eq.~(\ref{eq:brown}) is normalized to impact energy rather than to impactor diameter. $C_\lambda$ in Eq.~(\ref{eq:Clambda}) is proportional to $C_0$. Thus, $C_\lambda$ has the same power $A_E^{-9/10}$. This factor enters $\kappa_\mathrm{macro}$ linearly through Eq.~(\ref{eq:kappa_full}).

The total power of $A_E$ in $\kappa_\mathrm{macro}$ is $9/10-9/10=0$. This cancellation follows from using the same representative impactor density and velocity in the Brown energy-to-size conversion and in the lunar impact-energy calculation.

In this sense, we do not need the specific values for $\rho_\mathrm{imp}$ and $v_\mathrm{imp}$. However, their representative values are listed in Tables~\ref{tab:lunar_params} and \ref{tab:params}.

\subsubsection{Sensitivity of the absolute normalization}
\label{sec:sensitivity}

For the $q=2$ model, Eq.~(\ref{eq:normalization_scaling}) shows that $\eta$, $C_s$, and $C_i$ enter the normalization as $C_i(\eta/C_s)^{\beta/3}$. We therefore define the effective seismic-efficiency parameter
\begin{equation}
\eta_\mathrm{eff}\equiv\eta\left(\frac{C_i}{C_{i,0}}\right)^{3/\beta}\frac{C_{s,0}}{C_s},
\label{eq:etaeff}
\end{equation}
where $C_{i,0}=\pi/4$ and $C_{s,0}=4\pi$ are the fiducial values used in Appendix~\ref{sec:numerical}. This definition gives $\kappa_\mathrm{macro}\propto\eta_\mathrm{eff}^{\beta/3}$ when the other parameters are fixed. The parameter $\eta_\mathrm{eff}$ equals the physical seismic efficiency $\eta$ only when $C_i=C_{i,0}$ and $C_s=C_{s,0}$.

The fiducial result $1.7$~m$^2$~Myr$^{-1}$ exceeds the $0.28$~m$^2$~Myr$^{-1}$ estimate of \citet{Fassett2022} by a factor of about $6$. For $\beta=2.7$, matching the observational value requires
\begin{equation}
\eta_\mathrm{eff}\simeq10^{-4}\left(\frac{0.28}{1.7}\right)^{3/2.7}\simeq1.3\times10^{-5}.
\label{eq:etaeff_obs}
\end{equation}
If $C_i$ and $C_s$ are fixed at their fiducial values, this corresponds to $\eta\simeq1.3\times10^{-5}$, which lies within the broad range $\eta=10^{-5}$--$10^{-3}$ considered for seismic efficiency~\citep{Richardson2005}. If instead $\eta=10^{-4}$ is fixed, the same normalization can be obtained by changing $C_s$ or $C_i$. For example, keeping $C_i=C_{i,0}$ gives $C_s\simeq7.4C_{s,0}$, while keeping $C_s=C_{s,0}$ gives $C_i\simeq0.16C_{i,0}$. These examples only illustrate the degeneracy and are not independent estimates of the spatial factors.

The regolith density and thickness, vibration duration, impact-flux normalization, seismic-spreading factor, and impact-collection factor add further uncertainty to the absolute value. The comparison with observations should therefore be regarded as an order-of-magnitude consistency test. The more robust result is the crater-diameter exponent.

\subsection{Dependence on the effective spreading dimensionality}
\label{sec:comparison}

\begin{samepage}
The fiducial derivation assumes $V_\mathrm{eff}\propto D^2h_\mathrm{reg}$. For the generalized form of Eq.~(\ref{eq:Vq}), the vibration velocity scales as $v\propto\delta^{3/2}D^{-q/2}$ and the onset condition gives $\delta_\mathrm{min}\propto D^{q/3}$. Keeping $A_i=C_iD^2$, the same lower-limit-dominated integration yields
\begin{equation}
\kappa_\mathrm{macro}\propto D^{3-q\beta/3}.
\label{eq:qresult}
\end{equation}
Thus, for $\kappa_\mathrm{obs}\propto D^\xi$,
\begin{equation}
q=\frac{3(3-\xi)}{\beta}.
\label{eq:qinvert}
\end{equation}
Using $\xi=0.87$--$1.05$ from \citet{Fassett2022} and $\beta\simeq2.7$--$3.1$ gives $q\simeq1.9$--$2.4$, closer to shallow-layer spreading ($q=2$) than to homogeneous three-dimensional spreading ($q=3$).
\end{samepage}

For $q=3$, Eq.~(\ref{eq:qresult}) gives $\kappa_\mathrm{macro}\propto D^{3-\beta}$, with exponents from $0.3$ to $-0.1$ for $\beta\simeq2.7$--$3.1$, inconsistent with the near-linear lunar trend. The inferred $q$ should be regarded as an effective spreading exponent, not evidence for literal two-dimensional confinement. It can incorporate scattering, attenuation, leakage, and layer structure. Within the present model, the observations favor shallow-layer-like over homogeneous three-dimensional spreading.

\bibliography{CR-refs}
\bibliographystyle{aasjournalv7}

\end{document}